\documentclass[preprint,showpacs,preprintnumbers,amsmath,amssymb]{revtex4}

\usepackage{graphicx}

\begin{document}

\title{Triangular solution 
to general relativistic three-body problem for general masses}
\author{Kei Yamada}
%\email{}
\author{Hideki Asada} 
\email{asada@phys.hirosaki-u.ac.jp}
\affiliation{
Faculty of Science and Technology, Hirosaki University,
Hirosaki 036-8561, Japan} 

\date{\today}

\begin{abstract}

Continuing work initiated in an earlier publication 
[Ichita, Yamada and Asada, Phys. Rev. D {\bf 83}, 084026 (2011)], 
we reexamine the post-Newtonian effects on Lagrange's equilateral
triangular solution for the three-body problem.
For three finite masses, it is found that a triangular configuration 
satisfies the post-Newtonian equation of motion in general relativity, 
if and only if it has the relativistic corrections to each side length.
This post-Newtonian configuration for three finite masses 
is not always equilateral and it recovers 
previous results for the restricted three-body problem 
when one mass goes to zero.
For the same masses and angular velocity, 
the post-Newtonian triangular configuration is
always smaller than the Newtonian one. 

\end{abstract}


\maketitle

\section{Introduction}
One of classical problems in astronomy and physics is 
the three-body problem in the Newtonian gravity 
(e.g., \cite{Goldstein,Danby,Marchal}). 
The gravitational three-body problem is not integrable 
by the analytical method.
As particular solutions, however, 
Euler and Lagrange found a collinear solution and 
an equilateral triangular one, respectively.
The solutions for the restricted three-body problem, 
where one of three bodies is a test mass, are known as 
Lagrange points $L_1, L_2, L_3, L_4$ and $L_5$ \cite{Goldstein}. 
{\it Lagrange's equilateral triangular solution} has also 
a practical importance, since it is stable for some cases. 
Lagrange points $L_4$ and $L_5$ 
for the Sun-Jupiter system are stable and indeed 
the Trojan asteroids are located there. 
For the Sun-Earth system, asteroids were also found around $L_4$
by the recent observations \cite{CWV}.

Recently, Lagrange points have attracted renewed interests 
for relativistic astrophysics \cite{Krefetz,Maindl,SM,Schnittman,Asada,THA}, 
where they have discussed the relativistic corrections for Lagrange
points \cite{Krefetz,Maindl}, and the gravitational radiation 
reaction on $L_4$ and $L_5$ analytically \cite{Asada} 
and by numerical methods \cite{SM,Schnittman,THA}. 
It is currently important to reexamine Lagrange points in the framework 
of general relativity.
As a pioneering work, it was pointed out by Nordtvedt that 
the location of the triangular points is very sensitive 
to the ratio of the gravitational mass to the inertial one 
\cite{Nordtvedt}. 
Along this course, it might be important as a gravity experiment 
to discuss the three-body coupling terms in the post-Newtonian (PN) force,  
because some of the terms are proportional to a product of 
three masses as $M_1 \times M_2 \times M_3$. 
Such a triple product can appear only for relativistic three (or more) 
body systems but cannot for a relativistic compact binary 
nor a Newtonian three-body system. 

It was shown by Ichita et al., including the present authors, 
that a relativistic equilateral triangular
solution does not satisfy the equation of motion at the first
post-Newtonian (1PN) order except for two cases 
\cite{IYA}; 
(1) three finite masses are equal and 
(2) one mass is finite and the other two are zero. 
Hence, it is interesting to investigate
what happens at the 1PN level for three unequal finite masses in 
Lagrange's equilibrium configuration.
For the restricted three-body problem, on the other hand, 
Krefetz found a relativistic triangular solution by adding 
the corrections to the position of the third body \cite{Krefetz}. 
For three general finite masses, we shall look for a relativistic 
equilibrium solution that corresponds to Lagrange's equilateral 
triangular one.

Throughout this paper, we take the units of $G=c=1$. 

\section{Newtonian equilateral triangular solution}

\begin{figure}[h]
\begin{center}
\includegraphics[width=10cm]{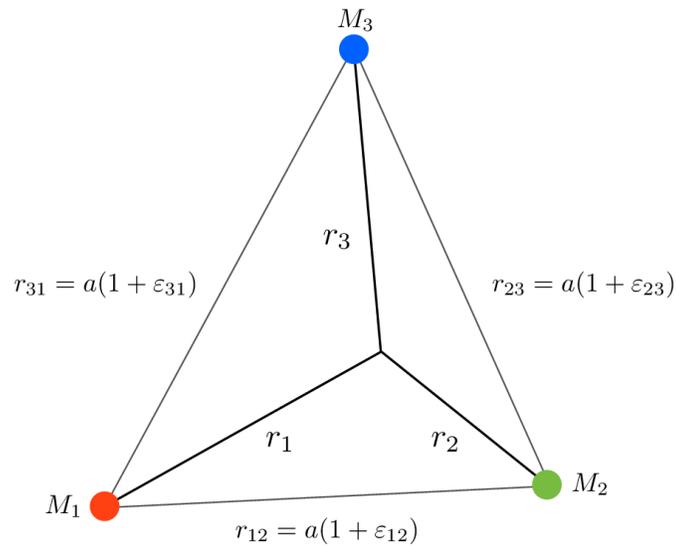}
\caption{PN triangular configuration. 
Each mass is located at one of the apexes.
$r_I \equiv |\boldsymbol{r}_I|~(I=1,2,3)$ denotes the orbital radius
of each body.
Each $\varepsilon_{IJ}$ denotes the relativistic correction to each
side length at the 1PN order.
In the equilateral case, $\varepsilon_{12}=\varepsilon_{23}=\varepsilon_{31}=0$,
namely, $r_{12}=r_{23}=r_{31}=a$.
}
\label{fig-tri-all}
\end{center}
\end{figure}

First, we consider the Newtonian gravity among three masses 
denoted as $M_I$ $(I=1, 2, 3)$ in a circular motion. 
The location of each mass is written as 
$\boldsymbol{r}_I$, where we choose the origin of the coordinates
as the common center of mass, so that 
\begin{equation}
M_1 \boldsymbol{r}_1
+ M_2 \boldsymbol{r}_2 
+ M_3 \boldsymbol{r}_3 = 0 . 
\end{equation}
We start by seeing whether 
the Newtonian equation of motion for each body can be satisfied 
if the configuration is an equilateral triangle. 
Let us put $r_{12} = r_{23} = r_{31} \equiv a$, 
where we define the relative position between masses as 
\begin{equation}
\boldsymbol{r}_{IJ} 
\equiv \boldsymbol{r}_I - \boldsymbol{r}_J , 
\end{equation}
and $r_{IJ} \equiv |\boldsymbol{r}_{IJ}|$ 
for $I, J = 1, 2, 3$. 
Then, the equation of motion for each mass becomes 
\begin{equation}
\frac{d \boldsymbol{r}_I}{dt^2} 
= - \frac{M}{a^3} \boldsymbol{r}_I , 
\label{N-EOM}
\end{equation}
where $M$ denotes the total mass $\sum_I M_I$. 
Therefore, it is possible that each body moves around 
the common center of mass with the same orbital period. 
Figure \ref{fig-tri-all} shows a triangular configuration for general masses.

Eq. (\ref{N-EOM}) gives 
\begin{equation}
\omega_{\text{N}}^2 = \frac{M}{a^3} , 
\label{omega2}
\end{equation}
where $\omega_{\text{N}}$ denotes the Newtonian angular velocity. 
The orbital radius $r_I \equiv |\boldsymbol{r}_I|$ 
of each body with respect to the common center of mass 
is obtained as \cite{Danby} 
\begin{align}
r_1 &= a \sqrt{\nu_2^2 + \nu_2\nu_3 + \nu_3^2} , 
\label{N-r1}
\\
r_2 &= a \sqrt{\nu_1^2 + \nu_1\nu_3 + \nu_3^2} ,
\label{N-r2}
\\
r_3 &= a \sqrt{\nu_1^2 + \nu_1\nu_2 + \nu_2^2} , 
\label{N-r3}
\end{align}
where we define the mass ratio as $\nu_I \equiv M_I/M$.

\section{post-Newtonian equilateral triangular solution}

Next, let us study the dominant part of general relativistic
effects on this solution.
Namely, we take account of the term at the 1PN order by employing 
the Einstein-Infeld-Hoffman (EIH) equation of motion 
in the standard PN coordinate as \cite{MTW,LL,AFH} 
\begin{align}
\frac{d \boldsymbol{v}_K}{dt} 
&= \sum_{A \neq K} \boldsymbol{r}_{AK} 
\frac{M_A}{r_{AK}^3} 
\left[
1 - 4 \sum_{B \neq K} \frac{M_B}{r_{BK}} 
- \sum_{C \neq A} \frac{M_C}{r_{CA}} 
\left( 1 - 
\frac{\boldsymbol{r}_{AK} \cdot \boldsymbol{r}_{CA}}
{2r_{CA}^2} \right) 
\right.
\nonumber\\
&~~~
\left. 
~~~~~~~~~~~~~~~~~~~~~
+ v_K^2 + 2v_A^2 - 4\boldsymbol{v}_A \cdot \boldsymbol{v}_K 
- \frac32 \left( 
\boldsymbol{v}_A \cdot \boldsymbol{n}_{AK} \right)^2 
\right]
\nonumber\\
&~~~
- \sum_{A \neq K} (\boldsymbol{v}_A - \boldsymbol{v}_K) 
\frac{M_A \boldsymbol{n}_{AK} \cdot 
(3 \boldsymbol{v}_A - 4 \boldsymbol{v}_K)}{r_{AK}^2} 
\nonumber\\
&~~~
+ \frac72 \sum_{A \neq K} \sum_{C \neq A} 
\boldsymbol{r}_{CA} 
\frac{M_A M_C}{r_{AK} r_{CA}^3} , 
\label{EIH-EOM}
\end{align}
where $\boldsymbol{v}_I$ denotes the velocity of each mass 
in an inertial frame 
and we define 
\begin{align}
\boldsymbol{n}_{IJ} \equiv
\frac{\boldsymbol{r}_{IJ}}{r_{IJ}} .
\end{align}
Note that Eq. (\ref{EIH-EOM}) for the EIH equation expresses the acceleration
of each mass,
where the force exerted on one mass is divided by the mass.
The PN force includes a product of three masses, 
whereas the acceleration by Eq. (\ref{EIH-EOM}) does that of two masses.

We consider three masses in a circular motion with the angular velocity
$\omega$, so that each $r_I$ can be a constant.
In addition, the common center of mass remains unchanged for the
equilateral triangular configuration as shown in \cite{IYA}.
Hence, the PN location $\boldsymbol{r}_{I}$ and orbital
radius $r_I$ of each body are unchanged from the Newtonian ones. 
As a consequence, 
the equation of motion for $M_1$ can be written as \cite{IYA}
\begin{align}
- \omega^2 \boldsymbol{r}_1 
&=
- \frac{M}{a^3}\boldsymbol{r}_1 
+ \boldsymbol{\delta}_{\text{EIH} 1} ,
\label{M1-EOM}
\end{align}
where $\boldsymbol{\delta}_{\text{EIH} 1}$ denotes 
the PN terms defined as
\begin{align}
\boldsymbol{\delta}_{\text{EIH} 1} &= 
\frac{1}{16}\frac {M^2}{a^3} \frac{1}{\sqrt{\nu_2^2 + \nu_2\nu_3 + \nu_3^2} }
\notag\\
&~~~
\times
\biggl\{\{16 (\nu_2^2 + \nu_2\nu_3 + \nu_3^2)
[3 - (\nu_1\nu_2 + \nu_2\nu_3 + \nu_3\nu_1)]
\notag\\
&~~~
+ 9\nu_2\nu_3[2(\nu_2 + \nu_3) + \nu_2^2 + 4\nu_2\nu_3 + \nu_3^2]\}
\boldsymbol{n}_{1}
\notag\\
&~~~
+ 3\sqrt{3}\nu_2\nu_3(\nu_2 - \nu_3)(5 - 3\nu_1)
\boldsymbol{n}_{\perp 1}
\biggr\} ,
\label{delta}
\end{align}
using $\boldsymbol{n}_{1} = \boldsymbol{r}_1/r_1$ and
$\boldsymbol{n}_{\perp 1} = \boldsymbol{v}_1/r_1\omega$ defined as the
unit normal vector to $\boldsymbol{r}_1$.
Eqs. (\ref{M1-EOM}) and (\ref{delta}) seem to disagree with 
Eqs. (31) and (32) in \cite{IYA}.
However, it is not the case.
This equality can be shown by noting that the angular velocity in the PN term, 
which appears at Eqs. (31) and (32) in \cite{IYA}, is equal to $M/a^3$.
One can obtain the equation of motion for $M_2$ and $M_3$ by cyclic
manipulations as $1 \to 2 \to 3 \to 1$.
These expressions show that the equilateral triangular solution is
present at 1PN order only for two cases; 
(1) three finite masses are equal and (2) one mass
is finite and the other two are zero \cite{Note}.

\section{post-Newtonian triangular solution for three finite masses}

For the restricted three-body problem, an inequilateral triangular
solution was investigated \cite{Krefetz,Maindl}.
Hence, for three finite masses, we study a PN triangular configuration.

Let us denote each side length of a PN triangle as
\begin{align}
r_{IJ} = a (1 + \varepsilon_{IJ}) ,
\label{ep}
\end{align}
where $\varepsilon_{IJ}$ denotes the non-dimensional correction 
at the 1PN order (see Fig. \ref{fig-tri-all}).
Here, if all the three corrections are equal 
(i.e. $\varepsilon_{12} = \varepsilon_{23} = \varepsilon_{31} = \varepsilon$), 
a PN configuration is still an equilateral triangle, 
though each side length is changed by a scale transformation 
as $a \to a(1 + \varepsilon)$.
Namely, one of the degrees of freedom for 
($\varepsilon_{12}$, $\varepsilon_{23}$, $\varepsilon_{31}$) 
corresponds to a scale transformation, and this is unphysical.
In other words, we take account of only the corrections 
which keep the size of the system but change its shape.
For its simplicity, we adopt the arithmetic mean of three side lengths 
in order to characterize the size of the system as
\begin{align}
\frac{r_{12} + r_{23} + r_{31}}{3} 
= a \biggl[ 1 + \frac{1}{3}
(\varepsilon_{12} + \varepsilon_{23} + \varepsilon_{31}) \biggr] ,
\label{add}
\end{align}
(see Appendix \ref{length} for possible choices 
of fixing the unphysical degree of freedom). 
This arithmetic mean of the PN triangle is chosen to be 
the same as a side length of 
the Newtonian equilateral triangle as
\begin{align}
 a \biggl[ 1 + \frac{1}{3}
(\varepsilon_{12} + \varepsilon_{23} + \varepsilon_{31}) \biggr]
= a ,
\end{align}
so that the degree of freedom for a scale transformation can be fixed.
Otherwise, a degree of freedom for a scale transformation would 
remain so that an ambiguity due to the similarity could enter our results.
Thus, we obtain a constraint on 
($\varepsilon_{12}$, $\varepsilon_{23}$, $\varepsilon_{31}$) as
\begin{align}
\varepsilon_{12} + \varepsilon_{23} + \varepsilon_{31} = 0 .
\label{constraint}
\end{align}
Hence, we look for the remaining two conditions for determining 
($\varepsilon_{12}$, $\varepsilon_{23}$, $\varepsilon_{31}$) 
in the following.

We assume a circular motion of each body, where the angular velocity 
of each mass is denoted as $\omega_I~(I=1,2,3)$.
At the 1PN order, the equation of motion for $M_1$ becomes
\begin{align}
- \omega_1^2 \boldsymbol{r}_1 &= 
M_2 \frac{\boldsymbol{r}_{21}}{r_{21}^3}
+ M_3 \frac{\boldsymbol{r}_{31}}{r_{31}^3} 
+ \boldsymbol{\delta}_{\text{EIH} 1}
\notag\\
&= 
- \frac{M}{a^3} \boldsymbol{r}_{1}
- \frac{3}{2} \frac{M}{a^2} \frac{1}{\sqrt{\nu_2^2 + \nu_2\nu_3 + \nu_3^2}}
\notag\\
&~~~
\times \{ [\nu_2(\nu_1 - \nu_2 - 1)\varepsilon_{12} 
+ \nu_3(\nu_1 - \nu_3 - 1)\varepsilon_{31}]\boldsymbol{n}_{1} 
+ \sqrt{3}\nu_2\nu_3
(\varepsilon_{12} - \varepsilon_{31})\boldsymbol{n}_{\perp 1} \} 
\notag\\
&~~~
+ \boldsymbol{\delta}_{\text{EIH} 1} + \mathrm{O}(\text{2PN}) .
\label{M1-EOM-C}
\end{align}
Note that each mass location $\boldsymbol{r}_I$ 
may be different from the Newtonian one, 
because the origin of the coordinates is chosen 
as the common center of mass in the 1PN approximation.
However, we can replace $\boldsymbol{r}_I$ with the Newtonian 
location of $M_1$ because of the following reason.
The two terms including $\boldsymbol{r}_I$
of Eq. (\ref{M1-EOM-C}) are expanded as 
\begin{align}
- \omega_1^2 \boldsymbol{r}_{1}
&=
- \omega_1^2 \boldsymbol{r}_{\text{N}1} 
- \omega_{\text{N}}^2 \boldsymbol{r}_{\text{PN}1} 
+ \mathrm{O}(\text{2PN}) ,
\label{pnterm1}
\\
- \frac{M}{a^3} \boldsymbol{r}_{1}
&=
- \frac{M}{a^3} \boldsymbol{r}_{\text{N}1} 
- \frac{M}{a^3} \boldsymbol{r}_{\text{PN}1} ,
\label{pnterm2}
\end{align}
respectively, where $\boldsymbol{r}_{\text{N}1}$ and
$\boldsymbol{r}_{\text{PN}1}$ denote the Newtonian location 
and the 1PN correction, respectively.
By using Eq. (\ref{omega2}), 
Eqs. (\ref{pnterm1}) and (\ref{pnterm2}) imply that
the 1PN corrections to $\boldsymbol{r}_I$ cancel out 
in Eq. (\ref{M1-EOM-C}).

Furthermore, $\boldsymbol{n}_{1}$ and $\boldsymbol{n}_{\perp 1}$ 
also have PN corrections. 
However, these corrections multiplied by $\varepsilon_{12}$ 
(or $\varepsilon_{31}$) make 2PN (or higher order) contributions 
in Eq. (\ref{M1-EOM-C})
and hence they can be neglected.
Also in $\boldsymbol{\delta}_{\text{EIH} 1}$, 
1PN corrections to $\boldsymbol{n}_{1}$ and $\boldsymbol{n}_{\perp 1}$ 
lead to 2PN, since they are multiplied by 1PN term as $M^2/a^3$.
We obtain the equation of motion for $M_2$ and $M_3$ by cyclic
manipulations as $1 \to 2 \to 3 \to 1$.

The PN equilibrium configurations can be present if and only if 
the following conditions (a) and (b) hold.
(a) Each mass has to satisfy the EIH equation of motion 
and (b) a triangular configuration does not change with time. 
Condition (a) is equivalent to 
(a') the coefficients of $\boldsymbol{n}_{\perp I}$ in the
equation of motion for each mass are zero:
\begin{align}
\varepsilon_{12} - \varepsilon_{31} 
- \frac{1}{8} \frac{M}{a} (\nu_2 - \nu_3)(5 - 3\nu_1) &= 0 ,
\label{condition1}
\\
\varepsilon_{23} - \varepsilon_{12} 
- \frac{1}{8} \frac{M}{a} (\nu_3 - \nu_1)(5 - 3\nu_2) &= 0 ,
\label{condition2}
\\
\varepsilon_{31} - \varepsilon_{23} 
- \frac{1}{8} \frac{M}{a} (\nu_1 - \nu_2)(5 - 3\nu_3) &= 0 .
\label{condition3}
\end{align}
Condition (b) is restated as 
(b') the angular velocity for each mass is the same in order to
keep the distance between masses unchanged:
\begin{align}
\omega_1^2 - \omega_2^2 &= 0 ,
\label{condition4-1}
\\
\omega_1^2 - \omega_3^2 &= 0 .
\label{condition5-1}
\end{align}
Eqs. (\ref{condition4-1}) and (\ref{condition5-1}) 
are rewritten as
\begin{align}
&~~~
\frac{3}{2}\frac{M}{a^3}
\frac{1}{\nu_2^2 + \nu_2\nu_3 + \nu_3^2}
[\nu_2(\nu_1 - \nu_2 - 1)\varepsilon_{12} 
+ \nu_3(\nu_1 - \nu_3 - 1)\varepsilon_{31}]
\notag\\
&~~~
- \frac{3}{2}\frac{M}{a^3}
\frac{1}{\nu_3^2 + \nu_3\nu_1 + \nu_1^2}
[\nu_3(\nu_2 - \nu_3 - 1)\varepsilon_{23} 
+ \nu_1(\nu_2 - \nu_1 - 1)\varepsilon_{12}]
\notag\\
&~~~
- \frac{M^2}{a^4}
\biggl\{\frac{9}{16}\frac{1}{\nu_2^2 + \nu_2\nu_3 + \nu_3^2}
\nu_2\nu_3
[2(\nu_2 + \nu_3) + \nu_2^2 + 4\nu_2\nu_3 + \nu_3^2] \biggr\} 
\notag\\
&~~~
+ \frac{M^2}{a^4} 
\biggl\{ \frac{9}{16}\frac{1}{\nu_3^2 + \nu_3\nu_1 + \nu_1^2}
\nu_3\nu_1
[2(\nu_3 + \nu_1) + \nu_3^2 + 4\nu_3\nu_1 + \nu_1^2] \biggr\} 
= 0 ,
\label{condition4}
\\
&~~~
\frac{3}{2}\frac{M}{a^3}
\frac{1}{\nu_2^2 + \nu_2\nu_3 + \nu_3^2}
[\nu_2(\nu_1 - \nu_2 - 1)\varepsilon_{12} 
+ \nu_3(\nu_1 - \nu_3 - 1)\varepsilon_{31}]
\notag\\
&~~~
- \frac{3}{2}\frac{M}{a^3}
\frac{1}{\nu_1^2 + \nu_1\nu_2 + \nu_2^2}
[\nu_1(\nu_3 - \nu_1 - 1)\varepsilon_{31} 
+ \nu_2(\nu_3 - \nu_2 - 1)\varepsilon_{23}]
\notag\\
&~~~
- \frac{M^2}{a^4} 
\biggl\{\frac{9}{16}\frac{1}{\nu_2^2 + \nu_2\nu_3 + \nu_3^2}
\nu_2\nu_3
[2(\nu_2 + \nu_3) + \nu_2^2 + 4\nu_2\nu_3 + \nu_3^2] \biggr\} 
\notag\\
&~~~
+ \frac{M^2}{a^4} 
\biggl\{\frac{9}{16}\frac{1}{\nu_1^2 + \nu_1\nu_2 + \nu_2^2}
\nu_1\nu_2
[2(\nu_1 + \nu_2) + \nu_1^2 + 4\nu_1\nu_2 + \nu_2^2] \biggr\} 
= 0 ,
\label{condition5}
\end{align}
respectively.
It seems that 
($\varepsilon_{12}$, $\varepsilon_{23}$, $\varepsilon_{31}$) 
do not always satisfy the above five conditions 
Eqs. (\ref{condition1}) - (\ref{condition5-1}) simultaneously.
However, the number of independent conditions turns out to be two.
The reason is as follows.
By eliminating $\varepsilon_{12}$ from 
Eqs. (\ref{condition1}) and (\ref{condition2}), 
we obtain Eq. (\ref{condition3}).
Moreover, the left-hand sides of Eqs. (\ref{condition4}) and (\ref{condition5}) 
always vanish, 
if and only if Eqs. (\ref{condition1}) and (\ref{condition2})
are satisfied.
These can be seen by direct calculations.

Thus, we obtain the expressions for 
($\varepsilon_{12}$, $\varepsilon_{23}$, $\varepsilon_{31}$) as
\begin{align}
\varepsilon_{12} &= \frac{1}{24} \frac{M}{a}
[(\nu_2 - \nu_3)(5 - 3\nu_1) - (\nu_3 - \nu_1)(5 - 3\nu_2)] ,
\label{d12}
\\
\varepsilon_{23} &= \frac{1}{24} \frac{M}{a}
[(\nu_3 - \nu_1)(5 - 3\nu_2) - (\nu_1 - \nu_2)(5 - 3\nu_3)] ,
\label{d23}
\\
\varepsilon_{31} &= \frac{1}{24} \frac{M}{a}
[(\nu_1 - \nu_2)(5 - 3\nu_3) - (\nu_2 - \nu_3)(5 - 3\nu_1)] ,
\label{d31}
\end{align}
which recover previous results for the restricted 
three-body problem \cite{Krefetz}. 

Substituting Eqs. (\ref{d12}) and (\ref{d31}) into Eq. (\ref{M1-EOM-C}),
we obtain the angular velocity as
\begin{align}
\omega_1 = \omega_{\text{N}} \biggl(1 + \frac{M}{a}\omega_{\text{PN}} \biggr) ,
\end{align}
where
\begin{align}
\omega_{\text{PN}} = 
- \frac{1}{16}[29 - 14(\nu_1\nu_2 + \nu_2\nu_3 + \nu_3\nu_1)] .
\label{omega-PN}
\end{align}
By using $\nu_1 + \nu_2 + \nu_3 = 1$, one can immediately show
$\omega_{\text{PN}} < 0$, so that we find $\omega < \omega_{\text{N}}$
for the same masses and $a$.
In other words, for the same masses and angular velocity, 
the PN triangular configuration is always smaller than the Newtonian one. 

Table \ref{table1} shows the relativistic corrections of the distance
between each body for Lagrange point $L_4$ ($L_5$) of Solar system. 
Here we choose $M_1$ and $M_2$ as the Sun and each planet, respectively.
In the case of the restricted three-body problem ($\nu_3 \to 0$), 
it is convenient to use Eqs. (\ref{condition1}) and
(\ref{condition2}) rather than Eqs. (\ref{d12}) - (\ref{d31}), 
because it is natural to change not $r_{12}$ but location of $M_3$. 
Eq. (\ref{condition2}) implies that the correction of distance between
each planet and Lagrange point $L_4$ ($L_5$) is 
approximately $5/16$ of Schwarzschild radius of the Sun. 
Hence, we obtain the same values of this correction for Earth
and Jupiter.
The similar corrections are mentioned also in the previous paper \cite{Maindl}.
The above PN effects, however, are so tiny that they could be neglected 
in the near-future measurements \cite{YA}.

It is interesting to extend this 1PN work to higher PN orders 
for the gravitational wave physics 
(see \cite{DJS, IFA, ABIQ, Itoh} 
for the equation of motion and compact binaries).

\section{Conclusion}

We reexamined the post-Newtonian effects on Lagrange's equilateral
triangular solution for three-body problem.
For three finite masses, it was found that a general triangular 
configuration satisfies the post-Newtonian equation of motion 
in general relativity, if and only if it has the relativistic
corrections to each side length. 
It was shown also that the post-Newtonian triangular configuration is
always smaller than the Newtonian one 
for the same masses and angular velocity.
Studying the correction to stability of this configuration is left as
future work.

\section{Acknowledgments}
This work was supported in part (K. Y.) 
by Japan Society for the Promotion of Science, 
Grant-in-Aid for JSPS Fellows, No. 24108.

\appendix

\section{Choices of fixing the unphysical degree of freedom}
\label{length}
In stead of the arithmetic mean of the three side lengths, 
one might wish to use the geometric mean of them or the triangular area.
Hence, let us mention briefly these cases.

The geometric mean of three side lengths by Eq. (\ref{ep}) is
written up to the 1PN order as 
\begin{align}
(r_{12} r_{23} r_{31})^{1/3}
= a \biggl[
1 + \frac{1}{3}
(\varepsilon_{12} + \varepsilon_{23} + \varepsilon_{31}) 
\biggr]
+ \mathrm{O}(\varepsilon^2) ,
\label{mul}
\end{align}
where $\mathrm{O}(\varepsilon^2)$ denotes the second order of 
($\varepsilon_{12}$, $\varepsilon_{23}$, $\varepsilon_{31}$), 
namely at the 2PN order.
This expression is identical with the arithmetic mean by Eq. (\ref{add}).
In addition, we can obtain the condition that these means are equal to 
a side length of the Newtonian equilateral triangle as
\begin{align}
\varepsilon_{12} + \varepsilon_{23} + \varepsilon_{31} = 0 .
\label{epsilon-total}
\end{align}

Next, we consider a PN triangular area.
A triangular area $S$ is given by Heron's formula as
\begin{align}
S = \sqrt{s (s - r_{12})(s - r_{23})(s - r_{31})} ,
\label{area}
\end{align}
where 
\begin{align}
s = \frac{r_{12} + r_{23} + r_{31}}{2} .
\end{align}
Hence, substitution of Eq. (\ref{ep}) into Eq. (\ref{area}) leads to
\begin{align}
S = \frac{\sqrt{3}}{4}a^2\biggl[ 1 + \frac{2}{3}
(\varepsilon_{12} + \varepsilon_{23} + \varepsilon_{31}) 
\biggr] 
+ \mathrm{O}(\varepsilon^2),
\end{align}
which corresponds to the triangular area 
of each side length Eq. (\ref{mul}).
Therefore, a PN triangular area is equal to 
a Newtonian equilateral triangular one if and only if
\begin{align}
\varepsilon_{12} + \varepsilon_{23} + \varepsilon_{31} = 0 .
\end{align}
This is identical with the condition Eq. (\ref{epsilon-total}). 

As a consequence, at 1PN level, the arithmetic mean of 
the three side lengths, the geometric mean of them, 
and the triangular area lead to the same characterizing 
the size of the system.

\begin{table}[h]
\caption{The corrections for Lagrange point $L_4$ ($L_5$) of Solar system.
Eqs. (\ref{condition1}) and (\ref{condition2}) are used for the
evaluation. 
Here, we choose $M_1$ and $M_2$ as the Sun and each planet, respectively.
Thus, $r_{12}=a(1+\varepsilon_{12})$ is the distance between the Sun
and each planet.
}
  \begin{center}
    \begin{tabular}{llll}
\hline
Planet & Sun-$L_4$ ($L_5$) [m] & Planet-$L_4$ ($L_5$) [m] \\
\hline 
Jupiter & $-0.353$ & -923  \\
Earth & $-0.00111$ & -923  \\
\hline
    \end{tabular}
  \end{center}
\label{table1}
\end{table}


\begin{thebibliography}{99}
\bibitem{Note}
Ref. \cite{IYA} suggests that if two of the masses are finite and equal, and
the third one is zero, an equilateral configuration satisfies the EIH
equation of motion.
However, the angular velocity of each mass is not the same 
in this case.
Note that this case is not an equilibrium configuration.
\end{thebibliography}
\end{document}